\documentclass[pra, twocolumn, showpacs, preprintnumbers, amsmath, amssymb, superscriptaddress, aps]{revtex4}
\usepackage{amssymb}
\usepackage{latexsym}
\usepackage[dvips]{graphicx}

\usepackage{epsfig}

\usepackage{dcolumn}
\usepackage{amsmath}
\usepackage{amsfonts}
\usepackage{bm}
\usepackage{color}

\newcommand{\be}{\begin{equation}}
\newcommand{\ee}{\end{equation}}
\newcommand{\bea}{\begin{eqnarray}}
\newcommand{\eea}{\end{eqnarray}}

\newcommand{\p}{\partial}
\newcommand{\s}{\sigma}

\newcommand{\la}{\langle}
\newcommand{\ra}{\rangle}
\newcommand{\rd}{\mbox{d}}
\newcommand{\ri}{\mbox{i}}
\newcommand{\re}{\mbox{e}}

\renewcommand{\vec}[1]{{\bm #1}}

\begin{document}
\title{Universality classes of order parameters composed of many body bound states}
\author{ A. M. Tsvelik}
\affiliation{Condensed Matter Physics and Materials Science Division, Brookhaven National Laboratory, Upton, NY 11973-5000, USA}
 \date{\today } 
\begin{abstract} 
 This theoretical paper discusses  microscopic models giving rise to special types of order in which  conduction electrons are bound together with localized spins to create composite order parameters. It is shown that composite order is related to the formation of a spin liquid  with gapped  excitations carrying quantum numbers which are a fraction of those of electron. These spin liquids are special in the sense that their formation necessarily involves  spin degrees of freedom of both  the conduction and the localized electrons and can be characterized by nonlocal order parameters. A detailed description  of such spin liquid states is presented with a special care given to a demonstration of  their   robustness  against local perturbations preserving  the Lie group symmetry and the translational  invariance. 
 
\end{abstract}

\pacs{71.10.Hf, 71.10.Pm, 71.27.+a}

\maketitle
\section{Introduction}

 This paper  puts forward a theoretical description  of composite order parameters (COPs). Such order parameters emerge as a result of  condensation of many-body bound states of conduction electrons with collective modes of interacting magnetic moments. On the formal level the COPs  are expressed as products of spin operators of  localized electrons and various density operators of conduction electrons. The latter ones may  include multiple products of  charge, spin and pair densities.  Being related to many body bound states a formation of the COP's requires  strong correlations  and their study will take us to hitherto unexplored areas of physics. The first example of such COP was found  in the  Kondo-Heisenberg chain model by Zohar and the author \cite{zohar}; it included a bound state of the staggered magnetization with the  pair density, an analogue of the Fulde-Ferell-Ovchinnikov-Larkin (FFLO) state, but created without  magnetic field. It also included the bound state  with a Charge Density wave with a wave vector  proportional to the total electron density (including the density of localized electrons). Although one dimensional models can support only quasi long range order, a real order is possible in arrays of chains provided one manages to couple the corresponding COPs. This may be a problem since they usually carry  finite wave vectors, so that  a coupling between COPs with different wave vectors is suppressed due to  the momentum conservation.  This suppression mechanism was invoked in \cite{berg} to explain   the exotic two-dimensional  superconductivity found in a layered compound La$_{1.875}$Ba$_{0.125}$CuO$_4$ \cite{tranquada},\cite{josephson}. It has been suggested that the  superconducting order parameters belong to the staggered pair density COPs and the COPs from neighboring  layers do not couple since their wave vectors are perpendicular to each other.  
  

 Although the concept of composite order is a general one, to achieve a better understanding we need to consider models  which allow reliable and controlled calculations. I suggest that  Kondo-Heisenberg models provide ideal platforms for these kind of studies. The core physics of \cite{zohar},\cite{FL} is the following. The Kondo-Heisenberg model brings together conduction electrons in the form of one-dimensional electron gas (1DEG) and antiferromagnetically correlated localized spins. Taken by themselves both electron and spin subsystems are quantum critical. In a quasi-one-dimensional setting this means that (i) the low energy modes are chiral, (ii) the spin and charge degrees of freedom of the 1DEGs  are separated. These two facts  suggest a possibility of a highly entangled state where right moving spinons  from the 1DEG pair to left moving ones from the antiferromagnet and {\it vice versa}.   As a result  two independent spin liquids are formed,  each one uniting  spin degrees of freedom of opposite chirality from the 1DEG and the spin chain;  the charge sector of the 1DEG is left gapless and is populated by the Goldstone modes.  Such state has a hidden order associated with pairing of spinons from different chains and hence is  robust with respect to local perturbations. The realization of such spin liquid is possible when the band filling of the 1DEG is far from 1/2 so that the Kondo exchange cannot generate backscattering. Then the  relevant coupling is  between the spin currents of opposite chirality from the  1DEG and the spin chain.   As I have said, the resulting spin liquid is a sum of two liquids formed by spinons of opposite chirality hence being mirror images of each other. The corresponding excitations carry fractional quantum numbers. Since  such pairing takes place not between electrons, but between the spinons, this process  cannot be treated perturbatively or via any kind of mean field  making it even more interesting.
 
  In \cite{zohar},\cite{FL} we studied the simplest version of the Kondo-Heisenberg (KH) model where the localized spins have magnitude 1/2 and there is one electronic chain per each spin one. In this paper I demonstrate that this is just one possibility out of many. One can construct entire universality classes of KH models corresponding to different representations of Lie groups with different topological orders, different gapped excitations and different COPs and generically a particular model may have several COPs. 
  
   Below I consider two types of models. Both of them describe arrays of one-dimensional Kondo-Heisenberg wires. In  the models of the first type the wires are arranged in  "cables", such that  each one-dimensional unit consists of a chain of localized spins $S=n/2$ surrounded by a bunch of $n$ conducting chains with incommensurate band fillings. In fact, to call this spin chain the Heisenberg one is an abuse of the term since the spin-spin interaction I consider includes higher powers of $({\bf S}_n{\bf S}_{n+1})$. It would be more appropriate to call it Generalized  Heisenberg chain, but I will not do it for the sake of brevity.  

 In the suggested construction the gapped fractionalized excitations are able to propagate only along  a single cable even when the chains are connected into arrays. 
I will  argue, however, that the three-dimensional coupling does not destroy  these excitations, although it creates their bound states  which carry  quantum numbers of electron and can propagate between the cables. A similar construction has been recently used in \cite{mudry} in the context of Fractional Quantum Hall effect. The models of another type are SU(N) generalizations of the Kondo-Heisenberg ladders considered in \cite{FL}.

 The paper is organized as follows. In Sections I-IV I will consider one dimensional models. In Section V I will discuss their arrays. In  Section II I  derive the continuum limit description for the both types of models mentioned above.  
This continuum description is given  by  integrable field theories whose spin sector is gapped and has fractionalized excitations.  For the cable model $n \geq 2$ their statistics is non-Abelian, for the SU(N)-symmetric ladder model it is Abelian.In Section III I will construct the composite order parameter operators  There is a separate universality class for each symmetry group representation. The construction can be easily generalized for nonunitary Lie groups.   
In Section IV it will be shown that the composite order and fractionalized excitations are robust with respect to group symmetry preserving perturbations around the integrable point. I will analyze in detail the  perturbations breaking  the symmetry between  the exchange couplings  and the perturbations driving  the spin chain  away from the criticality. 
In Section V I will consider physics of arrays of the KH cables.  

 The paper has Conclusions and Acknowledgements section and several Appendices. 

\section{The core models} 

 The core models of the present paper  are of two kinds. 

One type of the model called Kondo-Heisenberg cable (KHC)   consists  of  a critical antiferromagnetic spin S=n/2 Takhtajan-Babujian chain (TBC) coupled by an antiferromagnetic exchange interaction to $n$ conducting chains containing a one-dimensional electron gas (1DEG):
\bea
&& H = \sum_k \sum_{a=1}^n\epsilon_a(k) \psi^+_{k,a\s}\psi_{k,a\s} + \nonumber\\
&& \frac{1}{2} \sum_{k,q}J^{ab}\psi^+_{k+q,a\alpha}\vec\s_{\alpha\beta}\psi_{k,b\beta} {\bf S}_q +\nonumber\\
&&  J_H\sum_l {\cal P}_n\Big({\bf S}_l{\bf S}_{l+1}\Big), \label{model1}
\eea
where $\psi^+_a,\psi_a$ are creation and annihilation operators of the 1DEG on chains $a=1,...n$, $\s^b$ are the Pauli matrices, ${\bf S}_l$ is the spin $S=n/2$ operator on site $j$ and ${\bf S}_q$ is its Fourier transform.  ${\cal P}_n(x)$ is the polynomial of  degree $n$ whose exact form is fixed by the integrability requirements \cite{leon},\cite{hratch}. For instance, ${\cal P}_1(x) =x, {\cal P}_2(x) = x-x^2, $ {\it etc.} It is assumed that $J^{ab} << J_H$ and  the 1DEGs have band fillings incommensurate with the TBC: $|2k_{F,a}a_0 -\pi | \sim 1$. Under these assumptions one can formulate the low energy description of (\ref{model1}), taking into account that the backscattering processes between excitations in the TBC and the 1DEGs are suppressed by the above incommensurability.  The effective theory is valid for energies much smaller than both the average Fermi energy $\epsilon_{F,a}$ and the exchange interaction  $J_H$ of the model (\ref{model1}). 

  The reader should not remain under impression that the obtained results require a fine tuning of the spin sector to the integrable point. It will be shown in Section IV that they remain robust against those perturbations around the integrable point which preserve the translational and the  SU(2) symmetry.

Another core  model is a SU(N) symmetric generalization of the Kondo-Heisenberg chain:
\bea
&& H = \sum_k \sum_{a=1}^N\epsilon(k) \psi^+_{k,a}\psi_{k,a} +  \frac{1}{2} \sum_{k,q}J^{l}\psi^+_{k+q,a}\tau^l_{ab}\psi_{k,b} T^l_q +\nonumber\\
&&  J_H\sum_n (T^l_{n+1} T^l_{n+1}), \label{model2}
\eea
where $T^l, (l=1,...N^2-1)$ are generators of the su(N) algebra in the single box representation.

 The KHC model (\ref{model1}) is a one-dimensional  version of Spin Fermion (SF) model frequently used to study  violations of the Landau Fermi liquid theory in the vicinity of quantum critical points. As has been demonstrated in \cite{FL}, an array of KH chains can be used as a quasi 1D SF model. Here the electrons also interact with a critical insulating subsystem. However, the interacting regime I am going to study is different from what is usually assumed. In the standard treatment of the SF model  the quantum character of the spin fluctuations is not important, it is suggested that the quantum features are generated by the conduction electrons. In the quasi 1D version of the SF model considered here  this is not the case: the quantum nature of the spins is responsible for  creation of the spin gap and the formation of the spin liquid.

\subsection{Continuum limit of model (\ref{model1})}

 As usual I start with   the linearization of the spectrum of the 1DEG: 
\bea
\epsilon_a(k) \approx \pm v_{F,a}(k \mp k_{F,a}), \label{ep}
\eea 
and introduce the right and the left moving fermions $R$ and $L$: 
\bea
\psi_a(x) = \re^{-\ri k_{F ,a}x} R_a(x) + \re^{\ri k_{F,a} x}L_a(x).\label{psi}
\eea

 
 In the rest of my paper I will employ the formalism of non-Abelian bosonization most adequate for the task. Although this version of bosonization is not as widely known as the Abelian one, it has a venerable history and has been discussed in literature. The most recent review  can be found  in \cite{mudry}.  
 
The continuum limit of the TBC chain is described by the SU$_n$(2) Wess-Zumino-Novikov-Witten  (WZNW) model. This is a critical theory whose primary fields  transform in the spin $S \leq n/2$ representations of the SU(2) group. The excitations are gapless with linear spectrum $\omega = v_H|q|, ~~ v_H = \pi J_H/2$.  In the continuum limit the spin operators are approximated as \cite{affleck},\cite{ziman}: 
\bea
{ \bf S}_l = [{\bf j}_R(x) + {\bf j}_L(x)] + \ri (-1)^l \mbox{Tr}\Big[\vec\s(h-h^+)\Big] +..., \label{S}
\eea
 ($x=la_0$) where the dots stand for less relevant operators, $a_0$ is the lattice distance, $h$ is the WZNW SU(2) matrix field.  The current operators $j^a_L, j^a_R$ satisfy the SU$_n$(2) Kac-Moody algebra: 
\bea
[j_R^a(x), j_R^b(x')] = \ri\epsilon^{abc}j^c_R(x)\delta(x-x') + \frac{\ri n}{4\pi}\delta_{ab}\delta'(x-x'),
\eea
with the same commutation relations for the left currents $j^a_L$. The electron spin ${\bf F}_R = \frac{1}{2}\sum_{a=1}^n R^+_a\vec\s R_a, ~~ {\bf F}_L = \frac{1}{2}\sum_{a=1}^n L^+_a\vec\s L_a$  satisfy the same algebra. The remarkable fact is that the WZNW Hamiltonian describing the low energy part of the TBC can be expressed solely in terms of the currents:
\bea
H_{WZNW} =\frac{2\pi v_H}{n+2}\int \rd x \Big(:{\bf j}_R{\bf j}_R: + :{\bf j}_L{\bf j}_L:\Big), \label{TBCwz}
\eea
The double dots denote  normal ordering. 

 Due to the incommensurability of the Fermi wave vectors in the continuum limit the Kondo term in (\ref{model1})  is reduced to the interaction of the currents \cite{KZ}(see also \cite{embedding}):
 \bea
 V_{ex} = \frac{J_K}{2}\int \rd x ({\bf j}_R+{\bf j}_L)(R^+_a\vec\s R_a +L^+_a\vec\s L_a) \label{Vex}
\eea
where $J_K = J^{aa}$ (all diagonal elements are taken to be equal) At $v_{F,1} =v_{F,2}$ the sum of the electronic currents adds up to a single SU$_n$(2) current
\be
{\bf F}_R = \sum_{a=1}^n R^+_a\vec\s R_a, ~~{\bf F}_L = \sum_{a=1}^n L^+_a\vec\s L_a.
\ee
The further simplification comes from the fact that the relevant part of (\ref{Vex}) contains only products of the currents of different chirality so that the marginal interaction $
V_{marg} = J_K({\bf F}_R{\bf j}_R + {\bf F}_L{\bf j}_L)$ can be dropped as the first approximation.  Hence only the SU$_n$(2) part of the 1DEGs is involved in the interaction. 

 Below I will use the fact that the Hamiltonian of $n$ copies of spin 1/2 noninteracting fermions with identical Fermi velocities (I will assume this to simplify the calculations) can be written as a sum of the U(1) Gaussian model and SU$_2$(n) and SU$_n$(2) WZNW models \cite{embedding},\cite{QFT}. 
The resulting Hamiltonian is   
\bea
&& {\cal H}_{eff} = {\cal H}_{orb} + \frac{v_F}{2}[(\p_x\Theta_c)^2 +(\p_x\Phi_c)^2] + {\cal H}_{spin} \label{model3}\\
&& {\cal H}_{orb} = \frac{2\pi v_F}{n+2}\sum_{A=1}^{n^2-1}\Big( :I^A_R I^A_R: + : I^A_L I^A_L:\Big) \label{charge}\\
&& {\cal H}_{spin} = \frac{2\pi v_F}{n+2}\Big(:{\bf F}_R{\bf F}_R: + :{\bf F}_L{\bf F}_L:\Big) \label{Sp}
\eea
where $I^A~~, A=1,...n^2-1$ are SU$_2$(n) currents. $\Phi$ and $\Theta$ are mutually dual bosonic fields.  Both the charge part and model (\ref{charge}) are critical, the spectrum is linear: $\omega = v_F|k|$. 

 Now we will put the relevant part of (\ref{Vex}), (\ref{Sp}) and (\ref{TBCwz}) together and rearrange the terms in such a way to obtain two  commuting Hamiltonians: 
 \bea
&& {\cal H}_{spin} + V_{ex} + {\cal H}_{WZNW} = {\cal H}_s^{(Rl)} + {\cal H}_s^{(Lr)}\label{spin}\\
&& {\cal H}_s^{(Rl)} = \label{GN1}\\
&&\frac{2\pi v_F}{n+2}:{\bf F}_R{\bf F}_R: + \frac{2\pi v_H}{n+2}:{\bf j}_L{\bf j}_L: + J_K {\bf F}_R{\bf j}_L\nonumber\\
&& {\cal H}_s^{(Lr)} = \label{GN2}\\
&& \frac{2\pi v_F}{n+2}:{\bf F}_L{\bf F}_L: + \frac{2\pi v_H}{n+2}:{\bf j}_R{\bf j}_R: + J_K {\bf F}_L{\bf j}_R \nonumber
\eea
Here $v_F,v_H = \pi J_H/2$ are the Fermi velocity of the 1DEGs and the spinon velocity of the TBC respectively. 
This factorization of the spin sector into two part with one being a mirror image of another is the key feature of the KHC model (\ref{model1}) from which everything else will follow.  Such factorization can be easily generalized for any other Lie group symmetry besides SU(2) and any other representation of the spin operators.

Models (\ref{GN1},\ref{GN2}) are strongly interacting and integrable \cite{me}. These are WZNW models perturbed by a marginally relevant current -current interaction. Their Bethe ansatz solution  has many common features with the solution of the multi-channel Kondo model \cite{multi},\cite{multi1},\cite{multi2}  with the difference that in the  case when the spins are represented by a single impurity the spectral gaps cannot be formed.   At $J_K >0$ the spectrum consists of gapped non-Abelian solitons. Each soliton carries a zero mode of Z$_n$ parafermion. The further details are provided in Section IV B and  Appendix A. The  dispersion relations $E(k)_{Lr} = E(-k)_{Rl} =E(k)$   are (see Fig. 1)
\bea
 E(k) = k(v_H-v_F)/2 +\sqrt{k^2(v_F+v_H)^2/4 + \Delta^2}, \label{disp}
\eea
where  $
\Delta = \Lambda g\exp(- \pi/g), ~~ g = J_K/(v_F + v_H),$ with $\Lambda$ being the ultraviolet cut-off, is the spin gap. Hence they describe spin liquids. Since these models are mirror images of each other, under open boundary conditions the many-body wave functions of the two copies coincide at the boundaries. 

\begin{figure}[!htb]
\includegraphics[width=\linewidth,clip]{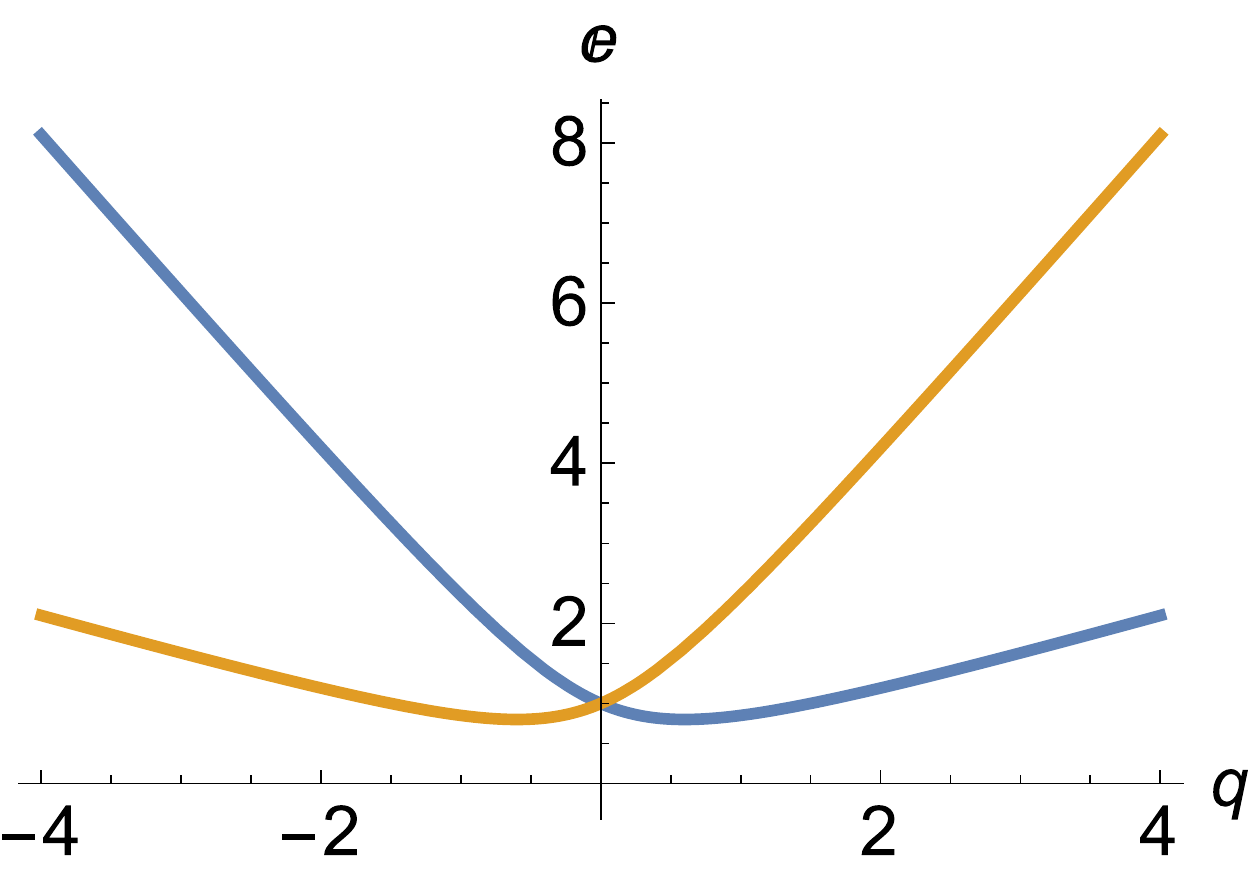}
\caption{The dispersion of the solitons  in the KH chain (\ref{disp}). $e= E/\Delta$, $q=k_x(v_Hv_F)^{1/2}/\Delta$ and  $v_F/v_H = 1/4$.}
 \label{Spec}
\end{figure}

\subsection{Continuum limit of model (\ref{model2})}

 The derivation here is very similar to the one given in the previous subsection. Therefore I will concentrate on differences. The first one is that the continuum limit of the SU(N) spin chain is now given by the SU$_1$(N) WZNW model. The SU(N) spin operators are expressed as 
\bea
T^l_n = [j^l_R(x) + j^l_L(x)] +  \sum_{q=1}^{N-1}\re^{2\pi nq/N}\mbox{Tr}\Big(\tau^l :h^q:\Big) +..., \label{S2}
\eea
where $h$ is the SU(N) matrix field of the SU$_1$(N) WZNW model and $\tau^l$ are generators of the su(N) algebra. The primary fields  $:h^q:$ are obtained  by fusion of the fundamental one their scaling dimensions are $d_q = q(N-q)/N$.

 The resulting continuum limit Hamiltonian density is 
 \bea
&& {\cal H}  = \frac{v_F}{2}[(\p_x\Theta_c)^2 +(\p_x\Phi_c)^2] + {\cal H}_s^{(Rl)} + {\cal H}_s^{(Lr)}\label{su(N)}\\
&& {\cal H}_s^{(Rl)} = \label{GN3}\\
&&\frac{2\pi v_F}{N+1}:F^l_R F^l_R: + \frac{2\pi v_H}{N+1}:j^lL j^l_L: + J_K^l F^l_R j^l_L\nonumber\\
&& {\cal H}_s^{(Lr)} = \label{GN4}\\
&& \frac{2\pi v_F}{N+1}:F^l_L F^l_L: + \frac{2\pi v_H}{N+1}:j^l_R j^l_R: + J_K^l F^l_L j^l_R \nonumber
\eea
Models (\ref{GN3},\ref{GN4}) can be written is a more familiar fermionic form. One can take advantage of the fact that SU$_1$(N) currents can be written in terms of fermionic bilinears and write the currents of the spin chain in terms of the auxiliary right- and left moving fermions $\rho,\lambda$.  The resulting model  constitutes the spin sector of the SU(N) Chiral Gross-Neveu model so that for (\ref{GN3}) we have  
\bea
&& {\cal H}_s^{(Rl)} =\label{fermionic}\\
&&  -\ri v_F R^+_{a}\p_x R_a  + \ri v_H\lambda^+_a\p_x\lambda_a + J_K^l(R^+\tau^lR)(\lambda^+\tau^l\lambda), \nonumber
\eea
where $a=1,2,... N$ with a similar expression with $x\rightarrow -x$ and $R$ replaced by $L$ and $\rho$ replaced by $\lambda$. 
 
When all coupling constants are equal $J_K^l =J_K$ models (\ref{GN3},\ref{GN4}) are integrable \cite{su(N)}, but now the spectrum contains $N-1$ branches of gapped excitations with spectral gaps $\Delta_j$:
\bea
&& \Delta_j = \Delta_1\frac{\sin(\pi j/N)}{\sin(\pi/N)}, ~~ j=1,2,...N-1,\\
&& \Delta_1 \sim  \exp[- \pi(v_F+v_H)/NJ_K] \nonumber
\eea
These excitations transform according to single column $j$-box irreducible representations of the SU(N) group. So they are fractional number particles. 
The spectrum of each branch is given by (\ref{disp}) with $\Delta$ replaced by $\Delta_j$. 

 When $J_K^l$ are different the picture remains qualitatively the same since the SU(N) symmetry is restored in the low energy limit \cite{balents}. 
 
\section{ Composite order parameters} 

In $D=1$ critical points are located at $T=0$ and there is only quasi long range order. Hence by order parameter (OP) operators I mean  the operators whose susceptibilities diverge at $T=0$. At $T=0$ their  correlation functions have a power law decay in space and time, at $T=0$ they decay exponentially with the correlation length $\sim 1/T$. In the core cable models such quasi long range order is expressed in term of OPs which include delocalized and  localized fermions -  composite order parameters (COPs). Ones 1D cables are arranged in a three dimensional array real long range order will be established. This will be discussed in more detail in Section V.

 \subsection{COPs in model (\ref{model1})}. 
 
 For simplicity sake  I will consider  the case $n=2$ in detail and discuss other cases briefly. 

 I will use the remarkable fact established in \cite{maldacena} that two noninteracting 1DEG with equal Fermi velocities can be described by the theory of eight Majorana fermions with O(8) symmetry. At the same time the SU$_2$(2) WZNW is equivalent to the theory of three noninteracting Majoranas. So the O(8) theory can be factorized into 5+3 Majoranas: $O_1(8) = O_1(5)\oplus O_1(3)$.

Remarkable properties of the SU$_2$(2) WZWN model has been first studied in \cite{FZ}. The reader can also find details in \cite{embedding},\cite{QFT}.  The SU$_2$(2) currents can be represented as products of Majorana fermions:
  \be
  j^a = \frac{\ri}{2}\epsilon^{abc}\kappa^b\kappa^c, ~~ F^a = \frac{\ri}{2}\epsilon^{abc}\chi^b\chi^c,
  \ee
  As a consequence the two independent Gross-Neveu models (\ref{GN1},\ref{GN2}) become the O(3) Gross-Neveu models of Majorana fermions:
  \bea
 && {\cal H}_s^{(Rl)} = -\frac{\ri v_F}{2}\chi^a_R\p_x\chi^a_R + \frac{\ri v_H}{2}\kappa^a_L\p_x\kappa_L^a +\nonumber\\
 && J_K\sum_{a>b}(\kappa^a_L\chi_R^a)(\kappa_L^b\chi_R^b),\label{Rl}\\
 && {\cal H}_s^{(Lr)} = \frac{\ri v_F}{2}\chi^a_L\p_x\chi^a_L - \frac{\ri v_H}{2}\kappa^a_R\p_x\kappa_R^a +\nonumber\\
 && J_K\sum_{a>b}(\kappa^a_R\chi_l^a)(\kappa_R^b\chi_l^b)\label{Lr}.
 \eea
The gapless sector given by the sum of (\ref{charge}) and the U(1) Gaussian model can be described as a model of 5 gapless Majoranas:
\bea
{\cal H}_{charge-orb} = \frac{\ri}{2}\sum_{a=1}^5(-\eta_R^a\p_x\eta_R^a + \eta_L^a\p_x\eta_L^a). 
\eea
For convenience we can group these fermions as follows: $\eta^{1,2}$ will correspond to fermionization of the charge sector, the other three $\eta$'s will describe the orbital sector. 

 At criticality the SU$_2$(2) WZNW model can also be represented as a sum of three critical quantum Ising models. This representation is particularly useful since the spin S=1/2 primary field (the matrix $h$) can be expressed in terms of order $\s_a$ and disorder $\mu_a$ parameter fields of the Ising models \cite{FZ}:
 \bea
&&  \hat h = \\
&& \hat\tau^0\s_1\s_2\s_3 +\ri(\hat\tau_1\mu_1\s_2\s_3 + \hat\tau^2\s_1\mu_2\s_3 + \hat\tau^3\s_1\s_2\mu_3). \nonumber
 \eea
Here $\tau^a, ~~ a=0,1...3$ are unit and Pauli matrices.

As is clear from (\ref{GN1},\ref{GN2}), the spectral gaps are generated by paring of Majoranas  of a given chirality from the 1DEG with their partners of opposite chirality from the TBC.  To clarify this it is instructive to do the  Hubbard-Stratonovich transformation for, for instance, model (\ref{Rl}). For $J_K>0$ the interaction is decoupled as  
 \bea
 J_K\sum_{a>b}(\kappa^a_L\chi_R^a)(\kappa_L^b\chi_R^b) \rightarrow \frac{\Delta^2}{2J_K} + \ri\Delta(\kappa^a_L\chi_R^a), \label{HS}
 \eea
 Integration over the fermions creates a double-well potential for field $\Delta$.  The minima of the potential correspond to degenerate vacua for the Majorana fermions where $\la (\kappa^a_L\chi_R^a)\ra \neq 0$.  As far as the operators of the original model (\ref{model1}) are concerned, the structure of the vacuum is more subtle since the local operators of this model are expressed not just in terms of  the Majorana fermion bilinears, but also in terms of Ising model operators (see Appendix B). A vacuum with one sign of $\Delta$ corresponds to the disordered phase of the Ising models where $\la\s^a\ra =0$, the other one corresponds to the ordered phase where $\la\s^a\ra \neq 0$ and may have any sign. Therefore the vacuum has a triple degeneracy. I will talk more about it in Section IV.

The important point is that since the Majoranas from the 1DEGs do not pair to each other, there are no order parameters   formed solely from the electronic operators  or spin operators. Instead, there are composite order parameters (COPs) whose correlation functions have a power law decay.

 As a preliminary step towards formulation of the COPs I will organize the fermions into Nambu spinors:
\bea
\Psi_{a\s} = \Big(
\begin{array}{c}
\psi_{\s,a}\\
\epsilon_{\s\s'}\psi^+_{\s',a}
\end{array}
\Big).
\eea
This reflects the orthogonal symmetry of the low energy sector. The spinor has 8 components; their quantum numbers include charge $q=\pm 1$, spin $\s = \pm 1$ and chain index $p= \pm 1$.  Products of the Nambu spinor components with the  right- and left chirality give rise to $8\times 8$ real matrix with 64 entries:
\bea
\Delta_{(q,p,\s),(q',p',\s')} = \bar r_{(q,p,\s)}l_{(q',p',\s')}
\eea
  Fusing it with the $4\times 4$ $h$-matrix spin field of the WZNW model one is left with the matrix COP containing 16 real entries: 
\bea
{\cal O}_{(q,p),(q',p')} = \bar r_{(q,p,\s)}h_{\s\s'}l_{(q',p',\s')}. \label{cop}
\eea
As it is  discussed in Appendix B,  this operator can be factorized into the part which condenses, acquiring a finite vacuum expectation value, and the part which fluctuates. The former one constitutes an amplitude of the fluctuating COP. The fluctuating part is a primary field of the critical O$_1$(5) theory with a scaling dimension 5/8. 
COP (\ref{cop}) contains charge density wave ($q= -q'$) and superconducting $(q= q')$ components. A given  matrix element carries the wave vector 
\be
Q_{(q,p),(q',p')}= qk_{F,p} -q'k_{F,p'} + \pi/a_0.
\ee

 Operators (\ref{cop}) constitute a reducible representation of the SO(5) group. This representation consists of an SO(5) scalar, vector, and antisymmetric tensor representations. To obtain the latter representations, one has to define five Dirac $\Gamma^a$ (a=1,...5) matrices, for instance,
 \bea
 && \Gamma^1 = \left(
 \begin{array}{cc}
 0 & \ri I\\
 -\ri I & 0
 \end{array}
 \right), ~~
 \Gamma^{2,3,4} = \left(
 \begin{array}{cc}
 \vec\s & 0\\
 0 & -\vec\s
 \end{array}
 \right), \nonumber\\
 && \Gamma^5 = \left(
 \begin{array}{cc}
 0 & -I\\
 - I & 0
 \end{array}
 \right),
 \eea
 where unit and Pauli matrices $I,\vec\s$ act on the chain indices. Then the ten SO(5) generators are defined as $\Gamma^{ab} = -\frac{\ri}{2}
[\Gamma^a,\Gamma^b]$. The corresponding COPs are defined as Tr${\cal O}$ (with wave vector $\pi/a_0$), Tr$\Gamma^a{\cal O}$ and Tr$\Gamma^{ab}{\cal O}$. Notice that besides the scalar COP which carries wave vector $\pi/a_0$, all others contain components with different wave vectors. The vector components with $a=1,5$ contain CDW order parameters with wave vectors $\pm [\pi/a_0 + 2(k_{F,1}+k_{F,-1})]$ (mod$[2\pi/a_0]$) corresponding to the total electron density (which includes the density of localized electrons). The same vector multiplet contains the $a=2$ component corresponding to the SC COP  with $\pi/a_0$ wave vector and $a=3,4$ components corresponding to CDWs with incommensurate wave vectors $\pm[k_{F,1}-k_{F,-1} +\pi/a_0]$.

 The above COPs (\ref{cop}) are not the only ones. One can make COPs by fusing products of fermionic bilinears with the higher spin primary fields of the SU$_n$(2) WZNW. For general $n$ these are the fields with spin $J \leq n/2$. For $n=2$ there are two such primary fields with $J=1/2, 1$ and hence there is only one extra operator: 
 \be
 \Phi_{ab} = \ri\kappa_R^a\kappa_L^b \label{J1}
\ee
As I have pointed out, the trace of this operator describes a smooth part of $({\bf S}_l{\bf S}_{l+1})$ lattice field. 
One can fuse (\ref{J1}) with either of the two operators 
\bea
&& (R^+_1\vec\s L_1)(R^+_{-1}\vec\s L_{-1}) \sim \re^{\ri\sqrt{4\pi}\Phi_c}\Big(\xi_R^a\xi_L^a - 3\eta_R^5\eta_L^5\Big),\nonumber\\
&& (R^+_1L_1)(R^+_{-1}R_{-1}) \sim \re^{\ri\sqrt{4\pi}\Phi_c}\Big(\xi_R^a\xi_L^a + \eta_R^5\eta_L^5\Big) \label{dens}
\eea
to get 
\bea
&& {\cal O}_{CDW}[2(k_{F,1} + k_{F,2})] = \label{cdw}\\
&& ({\bf S}_l{\bf S}_{l+1})(\psi^+_1\vec\s\psi_1)(\psi^+_{-1}\vec\s\psi_{-1})\re^{-2\ri(k_{F,1}+k_{F,2})x} \sim \re^{\ri\sqrt{4\pi}\Phi_c}, \nonumber
\eea
or with the product of two SC order parameter operators  
\bea
(R_1\s^yL_1)(R_{-1}\s^y L_{-1}) \sim \re^{\ri\sqrt{4\pi}\Theta_c}\Big(\xi_R^a\xi_L^a  + \eta_R^5\eta_L^5\Big),
\eea
to get a charge-4 ``bipairing'' operator
\bea
{\cal O}_{SC} =(\psi_1\s^y\psi_1)(\psi_{-1}\s^y\psi_{-1})({\bf S}_l{\bf S}_{l+1}) \sim \re^{\ri\sqrt{4\pi}\Theta_c}, \label{sc4}
\eea
which  existence of in four-leg ladders was discussed in \cite{chang},\cite{fradkin}. This  operator carries zero momentum.  To get other products one fuse, for instance
\bea
&& (R^+_1\vec\s L_{-1})(L^+_{-1}\vec\s R_{-1}), ~~(R^+_1L_{1})(L^+_{-1}R_{-1}) \sim \nonumber\\
&& \re^{\ri\sqrt{4\pi}\Phi_f}\Big(\xi_R^a\xi_L^a  + \eta_R^5\eta_L^5\Big). \label{other}
\eea
This operator carries zero charge and momentum $2(k_{F,1}-k_{F,-1})$.

All operators  (\ref{cdw},\ref{sc4},\ref{other}) have scaling dimension 1. They  are components of the  SO(5) symmetric tensor representation; in the Majorana language they are biproducts of right and left Majorana fermions $\eta^a_R\eta^b_L$.  

For higher $n$ one can fuse $2J$ fermionic bilinears with $J \leq n/2$-spin primary field of the spin chain to get operators with scaling dimension
\bea
d_J = 2\Big[J - \frac{J(J+1)}{n+2}\Big],
\eea
some of which will carry charge $Q= 4J$. However, for $J >1$ these operators have nonsingular susceptibilities. 

\subsection{COPs in model (\ref{model2})} 

 Below I will discuss only the case $N >2$, since the case $N=2$ is discussed at length in \cite{FL}.   

The primary fields of the SU$_1$(N) WZNW model are tensors in the antisymmetric representations described by a single column Young tableau with $q\leq N$ boxes.  They can be considered as products of fermion bilinears with the charge sector being factored out:
\bea
\Phi^{(q)} = \rho^+_{a_1}...\rho^+_{a_q}\lambda_{b_q}...\lambda_{b_1}\re^{q\sqrt{4\pi/N}\psi}, \label{primary}
\eea
where $\rho,\lambda$ are right and left moving Dirac fermions with velocity $v_H$ and $\psi$ is a real Gaussian bosonic field. Its correlation functions cancel the correlators of the charge field of the fermions. The scaling dimensions of (\ref{primary}) are 
\bea
d_q = \frac{q(N-q)}{N},
\eea
and they carry wave vectors $Q_q = \pm 2\pi q/Na_0$. 

 The COPs are SU(N) singlets and carry wave vectors $Q = (2k_F + 2\pi/Na_0)q$:
\bea
&& {\cal O}_q = (R^+_{a_1}L_{b_1})...(R^+_{a_q}L_{b_q})\Phi^{(q)}_{a_1,...a_q;b_1...b_q} = \nonumber\\
&& [(R^+_{a_1}\lambda_{a_1})(\rho^+_{b_1}L_{b_1})]^q
\re^{q\sqrt{4\pi}\psi} = A\re^{\ri q\sqrt{4\pi/N}\Phi_c}.\label{COPsuN}
\eea
The wave vector $(2k_F + 2\pi/Na_0)$ includes the density of localized and delocalized electrons in agreement with Oshikawa theorem \cite{oshikawa}. Ones the spin gaps are formed the amplitude $A$ is finite. The scaling dimensions are 
\be
d_q = q^2/N,
\ee
Notice that for $N>2$ the COPs are of the charge density wave type and does not include superconducting ones.  

\section{Robustness against perturbations}

 The spin liquid states described above represent only a part of the Hilbert space of the original models (\ref{model1},\ref{model2}). The rest of it belongs to gapless excitations. Hence the current models describe  conducting states. 
 Nevertheless since the spin sector is decoupled from the gapless modes (the charge and orbital ones for (\ref{model1}) and the charge one for (\ref{model2})) it is instructive to find out how robust are its fractionalized excitations  against various perturbations. Below I will consider several perturbations concentrating mostly on model (\ref{model1}) and show that the fractionalized gapped excitations are robust against  perturbations which do not violate the SU(2) (for model (\ref{model1})) and the SU(N) (for model (\ref{model2})) symmetry of the spin chain and do not break the translational invariance. 
 
  Such perturbations fall into several  categories which will be considered below. First, there are electron-electron interactions of the band electrons. Away from half filling they generate only current-current interactions. In Subsection A it will be demonstrated that such interactions together with current-current interactions in the  spin sector will  need to exceed some critical value to radically modify the spin liquid state. Second, there are perturbations in the spin chain which would destroy the SU$_n$(2) critical point of an isolated chain. They will be analysed in Subsection B. If not too strong such perturbations are ineffective since ones the spin liquid is formed its stability is protected by the spectral gap. Third, there are perturbations   corresponding to channel anisotropy $J^{11} \neq J^{22}$ which will be  discussed in Subsection C. At last, there is external magnetic field, but the spin liquid is protected against it by the spin gap. These four  categories exhaust the list of the symmetry preserving perturbations.
   
  \subsection{Electron-electron interactions of the band electrons}

 It is instructive to find out whether the gapped state described in the previous Section can be adiabatically connected to a topologically trivial state of decoupled band electrons  and a gapped TBC. To show this I introduce a deformation of the original model adding to it the additional interaction
\bea
V = \gamma({\bf F}_R{\bf F}_L + {\bf j}_R{\bf j}_L),
\eea
and consider a trajectory in the $J_K-\gamma$ plane from $(J_K,0)$ to $(0,\gamma)$. Since the charge-orbital sector remains decoupled the trajectory lies entirely inside of the spin sector which remains gapped except, as we will see, at one critical point separating the two phases.  One of those is  the  phase of interest and the other one is  phase where TBC and 1DEGs are disconnected. The spin excitations  are gapped; at   $J_K=0$ and $\gamma >0$ both the band electrons and the TBC are perturbed by the marginally relevant products of the currents. These are  integrable perturbations of  the same kind as in (\ref{GN1},\ref{GN2}); they generate  spectral gaps. For the spin chain there is also  OP local in the spin operators:
\be
{\cal O} = \la ({\bf S}_j{\bf S}_{j+1})\ra \sim \la \mbox{Tr}h \mbox{Tr}h^+\ra,
\ee
 which describes a spontaneously generated deviation from the integrable point. On the other hand,  the phase $\gamma =0$ has no local OPs, there is only a quasi long range order (see Section III). As we will see, the two phases are separated by a quantum critical point.

For simplicity I set $v_F = v_H$. Let us introduce new operators
\be
{\bf J} = {\bf F} +{\bf j}, ~~ {\bf K}= {\bf F}-{\bf j}.
\ee
The operators ${\bf J}_{R,L}$ are SU$_{2n}$(2) Kac-Moody currents. Then the total interaction becomes 
\be
V_{ex} + V = \frac{1}{2}(J_K+\gamma){\bf J}_R{\bf J}_L + \frac{1}{2}(\gamma-J_K){\bf K}_R{\bf K}_L
\ee
The  part of the Hamiltonian describing the critical point can be represented as the sum of the SU$_{2n}$(2) WZNW and the SU$_n$(2)$\times$SU$_{n}$(2)$/$SU$_{2n}$(2) coset theory. At $\gamma =J_K$ the product of the ${\bf K}$-operators vanishes and the latter theory decouples and becomes critical; so the entire theory has a critical point. At this point only SU$_{2n}$(2) part of the spin Hilbert space is gapped, the remaining SU$_n$(2)$\times$SU$_{n}$(2)$/$SU$_{2n}$(2) one is gapless. Hence the phase with small $\gamma$ is separated from the topologically trivial phase with $J_K =0$ by a quantum critical point described by the SU$_n$(2)$\times$SU$_{n}$(2)$/$SU$_{2n}$(2) coset theory. 

 In my opinion it is possible that the gapped spin state described above is topologically nontrivial. Indeed, it has nonlocal OP of the string type and is likely to have zero modes located on a boundary with the topologically trivial phase $\gamma > J_K$. However, in order to determine a place of this model in the general classification of topological phases \cite{Fid}, I have to consider the edge zero energy modes. I leave this problem for future studies.

 \subsection{Deviations from the SU$_n$(2) critical point}
 
  In this Subsection I demonstrate that the deviations of the spin chain from the TBC integrable point do not confine the non-Abelian massive excitations. For simplicity I do it for the $n=2$ KHC model. If the perturbation is not too strong it just creates bound states of the non-Abelian solitons, but these particles still   remain in the spectrum.

 I start with the unperturbed model for $n=2$. The spin sector is described by a sum of two copies of the O(3) Gross-Neveu model (\ref{Rl},\ref{Lr}).  The exact solution of the O(3) GN model was first found in \cite{susy} as a particular limit of the supersymmetric sine-Gordon model and was  later analyzed in detail in \cite{fendley},\cite{suzuki}. The reader can find an excellent and pedagogical analysis of (1+1)-dimensional supersymmetric theories in a recent paper by Mussardo \cite{mussardo}. As I have stated above (see the text around (\ref{disp})), the excitations are  massive and non-Abelian.  Their nature can be visualized with a help of Hubbard-Stratonovich transformation (\ref{HS}).  Then, as I have mentioned above, the integration over the Majorana fermions creates a double-well potential for field $\Delta$. The Majorana fermions have zero energy modes on the kinks of $\Delta$-field; the kinks with attached zero modes constitute  excitations of the O(3) GN model, so called Bohomol'nyi-Prasad-Sommerfield (BPS) solitons \cite{fendley}. A multi-kink state is a highly entangled one and cannot be factorized into a product of states even when the kinks are far from each other. This becomes clear when one considers a Hilbert space of Majorana zero modes. These  modes $\gamma_a$ obey Clifford algebra 
\be
\{\gamma_a,\gamma_b\} = \delta_{ab},
\ee
and the Hilbert space of  $N$ kinks have $2^{[3N/2]}$ states.  

 According to the exact solution \cite{susy},\cite{fendley},\cite{suzuki}  the excitation spectrum does not contain vector particles. Hence the Majorana fermions themselves do not survive as coherent excitations; this fact is important for the survival of the fractionalized particles. 

 For the following analysis it will be convenient to use the relativistic parameterization of the soliton spectrum (\ref{disp}):  
  \bea
  && E_p = \frac{\Delta}{2}\Big(\sqrt{v_F/v_H}\re^{p\theta} + \sqrt{v_H/v_F}\re^{-p\theta}\Big), \nonumber\\
  && P = \Delta (v_Hv_F)^{-1/2}\sinh\theta, \label{param}
  \eea
  where parameter $\theta$ is called rapidity. Different signs correspond to different copies of the O(3) GN model: $p=+$ for  (\ref{Rl}) and $p=-$ for (\ref{Lr}).  
  
  In description of the excitations I will follow \cite{dunning}. The ground state of a single O(3) GN model is triple degenerate. The excitations are solitons interpolating between differet vacua.  A soliton with rapidity $\theta$ interpolating between the vacua $a$ and $b$ is created  by operator $K^{\s}_{ab}(p,\theta)$ with $\s =\pm 1/2$ for soliton and antisoliton ($s^z = \pm 1/2$) respectively.  The vacuum indices $a,b$ take values 0, 1/2 and 1 with $|a-b| =1/2$. The latter restriction is responsible for the fact that a multisoliton state cannot be disentangled into a product of single-particle states even if solitons are far from each other. Multi-soliton state of a given model with total spin projection $S^z = \sum_j \s_j$ is given by 
\bea
|K^{\s_1}_{a_0a_1}(p,\theta_1)K^{\s_2}_{a_1a_2}(p,\theta_2)...K^{\s_N}_{a_{N-1}a_N}(p,\theta_N)|0_{a_N}\ra,
\eea
where $\theta_1> \theta_2>...>\theta_N $ for an $in$ and $\theta_1< \theta_2<...<\theta_N $ for an $out$ state. The 2-particle scattering process 
\bea
K^{\s_1}_{ab}(p,\theta_1) + K^{\s_2}_{bc}(p,\theta_2) \rightarrow K^{\s_2'}_{ad}(p,\theta_2) + K^{\s_1'}_{dc}(p,\theta_1),
\eea 
 is described by the scattering matrix 
\bea
S_{SUSY}\left(
\begin{array}{cc}
a & d\\
b & c
\end{array} | \theta_1-\theta_2 \right) \times
S_{\s_1,\s_2}^{\s_1',\s_2'}(\theta_1-\theta_2),
\eea
where $S_{SUSY}$ is described in \cite{dunning} and the other  $S$-matrix is the one of the SU(2) Thirring model:
\bea
&& S_{\s_1,\s_2}^{\s_1',\s_2'}(\theta) = -S_0(\theta)\frac{\Big( \theta \delta_{\s_1,\s_1'}\delta_{\s_2,\s_2'} + \ri\pi\delta_{\s_1,\s_2'}\delta_{\s_2,\s_1'}\Big)}{\theta + \ri\pi},\nonumber\\
&& S_0(\theta) = \frac{\Gamma(1/2-\theta/2\pi)\Gamma(1 + \ri\theta/2\pi)}{\Gamma(1/2+\theta/2\pi)\Gamma(1 - \ri\theta/2\pi)}
\eea
 
  The relevant operators of the SU$_2$(2) WZNW model include spin $S=1/2, 1$ primary fields and the product of the left and right currents. As is obvious from (\ref{S}), the $S=1/2$ operator breaks the translational invariance. If we do not allow this, the most relevant perturbation is the $S=1$ primary field which is local in the Majorana fermions:
  \be
  V_{pert} = \ri m\kappa_R^a\kappa_L^a \label{pert}
  \ee
  For $J_K =0$ this perturbation would lead to a confinement of the fractionalized excitations of the TBC \cite{tsvelik90},\cite{shelton}. However, as I am going to show, for finite $J_K > 0$ this is no longer the case provided $|m| << \Delta$.

For the following we will need to obtain some information about matrix elements  of the perturbing operator (\ref{pert}). Leaving a complete calculation for the future, I will just establish the properties necessary to resolve the problem of confinement. This can be done on the basis of Lorentz invariance and crossing symmetry. 

From the exact solution we know that the Majoranas are not coherent particles. Hence operator $\kappa_R$ ($\kappa_L$) has matrix elements between a vacuum and states of even number of solitons of model (\ref{Lr})(respectively of (\ref{Rl})). The minimal matrix elements corresponding to annihilation of two solitons are 
\bea
&& \la 0_a |\kappa_R^l(\tau,x)|K_{ab}^{\s_1}(-,\theta_1)K_{ba}^{\s_2}(-,\theta_2)|0_a\ra = \nonumber\\
&& \exp\Big\{-\tau [E_{Lr}(\theta_1)+E_{Lr}(\theta_2)] - \ri x[P(\theta_1)+P(\theta_2)]\Big\}\times\nonumber\\
&& \Delta^{1/2}\re^{(\theta_1+\theta_2)/4}g_{a}(\theta_1-\theta_2)C^{l}_{\s_1\s_2},\\
&& \la 0_a |\kappa_L^l(\tau,x)|K_{ab}^{\s_1}(+,\theta_1)K_{ba}^{\s_2}(+,\theta_2)|0_a\ra = \nonumber\\
&& \exp\Big\{-\tau [E_{Rl}(\theta_1)+E_{Rl}(\theta_2)] - \ri x[P(\theta_1)+P(\theta_2)]\Big\}\times\nonumber\\
&& \Delta^{1/2}\re^{-(\theta_1+\theta_2)/4}g_{a}(\theta_1-\theta_2)C^{l}_{\s_1\s_2}
\eea
where $C$ is the Klebsh-Gordon factor and $g_{a}(\theta)$ is a dimensionless function to be determined. This form is dictated by the fact that (i) $\kappa^l$ has spin 1 under the SU(2) group and the solitons have spin 1/2, (ii)  $\kappa_{R,L}$ are components of a spinor, that is they have Lorentz spin $\pm 1/2$. The latter fact explains the presence of the exponential factors:  under a Lorentz boost $\theta_i \rightarrow \theta_i + \alpha$ the matrix elements must acquire a factor $\re^{\pm\alpha/2}$. 

 We can extract more specific information about the matrix elements from the crossing symmetry. It  allows one to extract  another matrix element:
\bea
&& \la 0_a| K_{ab}^{-\s_1}(p;\theta_2)|\kappa_R^l(0,0)|K_{ab}^{\s_2}(p;\theta_1)|0_a\ra = \nonumber\\
&& \Delta^{1/2}\ri\re^{ p(\theta_1+\theta_2)/4}g_{a}(\ri\pi -\theta_1+\theta_2)C^{l}_{\s_1\s_2},
\eea
As we shall see, the issue of the soliton confinement is decided by the behavior of this matrix element at $\theta_1 \rightarrow \theta_2$. The solitons are confined if the function $g(\theta)$ has  a pole at $\theta =\ri\pi$. In that case the effective potential between the kinks grows with distance (see below). However, according to the general theorem (see, for instance, \cite{mussardo}) at the pole we have 
\bea
g_a(\ri\pi -\theta) \sim \frac{\la 0_a|\kappa_{R,L}^l|0_a\ra}{\theta}.
\eea
 and the residue is zero since $\kappa_{R,L}$ are fermion operators and cannot have a nonzero vacuum average. This conclusion is also supported by the semiclassical calculation for the supersymmetric sine-Gordon model done in \cite{mussardo} (see Eqs.(43,44)) which gives an explicit expression for $g_a(\theta)$.  

 Now we can use all this accumulated information to write down the Schr\"odinger equation for two solitons belonging to the sectors with  different parity. Their wave function is 
\bea
B_{\s_1,\s_2}\int \rd\theta_1\rd\theta_2 \Psi_{ab;cd}(\theta_1,\theta_2)K_{ab}^{\s_1}(+;\theta_1)K_{cd}^{\s_2}(-;\theta_2)|0_b,0_d\ra\nonumber
\eea
Acting on this state by (\ref{pert}) we create a two-soliton state  plus multi-soliton states. Since the latter ones lay higher in energy we can  neglect them when $m << \Delta$.


 In the reference frame with the total zero momentum we have 
 \begin{widetext}
 \bea
 \Big(-E + 2M\cosh\theta\Big)\Psi(\theta,-\theta) + (m\Delta/M)\hat P_{S=1}\int \frac{\rd u}{\cosh u} \re^{(\theta +u)/2}|g(\ri\pi +\theta-u)|^2\Psi(u,-u) =0,
 \eea
 \end{widetext}
 where $M = \Delta(v_H+v_F)/\sqrt{v_Hv_F}$ and $\hat P_{S=1}$ is a projector to spin $S=1$ space which sets the spins of the solitons  into a triplet configuration. 
 
 For $m << M$ one can expand the kernels in small rapidities and obtain the Schr\"odinger equation:
 \bea
 && \Big[-E +2M - \frac{1}{4M}\frac{\p^2}{\p x^2} + V(x)\hat P_{S=1}\Big]\tilde\Psi(x) =0, \nonumber\\
 && V(x) =(m\Delta/M) \int_{-\infty}^{\infty} \frac{\rd u}{2\pi}  \re^{2\ri M u x}|g(u +\ri \pi)|^2, \nonumber\\
 && \tilde\Psi(x) = \int_{-\infty}^{\infty} \frac{\rd u}{2\pi}  \re^{2\ri M u x}\Psi(u,-u).
 \eea
 The potential $V(x)$ decays at large distances unless function $g(\theta)$ has  a pole at $\ri\pi$. As we have already established, there is no pole.  For $m<0$ the potential is attractive and there is at least one  bound state below the two-particle continuum. Bound states do not kill the fractionalized excitations, they remain in the spectrum. The topologically nontrivial state survives.

\subsection{Asymmetry of the Kondo exchange}

 The asymmetry of the Kondo couplings is a marginally irrelevant perturbation which dies out under renormalization. This is the case for both types of models. The renormalization group dynamics of the KHC model (\ref{model1}) is identical to the one of the Kondo impurity. For the impurity problem it has been well known from the late 70-ties (see \cite{blandin}) that the stable exchange configuration is the one when the impurity spin is completely screened. This means that  when the impurity spin $S$ interacts with several screening channels with different exchange integrals, the renormalization selects $2S$ channels with strongest couplings  which become identical under the RG flow and suppresses all weaker ones. Likewise, the SU(N) symmetry is restored in strong coupling limit for model (\ref{model2}), as was shown in \cite{balents}.   
 
\section{The Kondo-Heisenberg arrays}

 In this Section I will discuss a generalization of  the "wire construction" of Kondo-Heisenberg arrays developed  in my previous publication \cite{FL}. Namely, I briefly consider an array of parallel  KHC models connected by interchain electron tunneling and exchange interactions. 

As it was discussed in \cite{FL}, the effect of these interactions is twofold. First, they couple the COPs which eventually leads to a real long range order. Second, the interchain tunneling and exchange create  coherent excitations. In particular, the tunneling create bound states of holons and spinons (quasiparticles) whose dispersion is located inside of the spinon gap. When the tunneling matrix elements  are sufficiently large (of order of the spinon gap) the quasiparticle dispersion crosses the chemical potential and a Fermi surface appears in the form of electron and hole   pockets \cite{FL} and formation of Fermi liquid. Likewise, the interchain exchange interaction leads to creation of bound states from fractionalized spin excitations which can propagate in the bulk. The fractionalized particles themselves remain confined to the chains, at least in the model I consider. I will not discuss these subjects further not to  distract attention from the main subject of this paper which is composite order. Instead I will consruct several possible Ginzburg-Landau (GL)  functionals for the COPs. 

 Model (\ref{model2}) provides the simplest example due to the simplicity of the order parameters (\ref{COPsuN}). Here they are just bosonic exponents with U(1) symmetry. Hence the GL Hamiltonian  for the array of (\ref{model2}) cables is 
\bea
&& H =\sum_r\int \rd x\Big\{ \frac{v_F}{2} [(\p_x\Theta_r)^2 + (\p_x\Phi_r)^2] + \nonumber\\
&& \sum_q \sum_{r'}J_{r,r'}^{(q)}\cos[q\sqrt{4\pi/N}(\Phi_r - \Phi_r')]\Big\},
\eea
where indices $r,r'$ mark positions of different chains. As was noticed in \cite{FL}, the interchain couplings $J$ are generated not just by the electron hopping, but also by the interchain spin exchange. This feature may lead to some interesting consequences as far as the ordering is concerned. For example, due to the composite nature of the OP it will not be so easy to pin the phase by disorder. Indeed, the pinning operator must simultaneosly act on the spins located on the central chains and on the electrons located on their own chains.  

 For model (\ref{model2}) the situation is reacher due to the presence of the orbital degrees of freedom related to spatial position of the 1DEGs around the central spin chain. Hence the coupling of the cables will in general break  all symmetries except of the U(1) charge. 

 It would be too tedious to discuss here all possible Ginzburg-Landau theories. I will just discuss one which is sufficiently exotic and interesting, namely, the theory of charge-4 ``biparing'' superconductivity related to condensation of COP (\ref{sc4}). This is the only relevant COP which has zero momentum. To suppress coupling between all  other COPs  one should arrange the cables in such a way that their mutual positions prevent a coupling of COPs with finite wave vectors. An example of such arrangement is a pyroclore lattice where all chains intersect each other at finite angles. The resulting GL Hamiltonian is 
\bea
&& H = \sum_r\int \rd x\Big\{ \frac{v_F}{2} [(\p_x\Theta_r)^2 + (\p_x\Phi_r)^2] + \nonumber\\
&& \sum_{r'}J_{r,r'}\cos[\sqrt{4\pi}(\Theta_r - \Theta_{r'})]\Big\},
\eea
where $\Theta_r$ is $4e$ charged phase field on chain $r$.

 \subsection{Influence of perturbations on the ordering}
 
  The presence of various perturbations can generate additional couplings between COPs from different cables. For instance, since the fusion of operator  (\ref{pert}) with four  conduction electron operators gives rise to COPs (\ref{cdw},\ref{sc4}), the deviation from the quantum critical point of the spin model   helps to establish an interchain coupling of the quartic operators. In the presence of such perturbation these COPs couple by the  interchain hopping alone. On the other hand the perturbations which break the translational invariance of the spin chain generate operators Tr$(h+h^+)$ or iTr[$\vec\s(h-h^+)]$ and hence through (\ref{cop}) generate  a coupling between the conventional CDW and superconducting order parameters. Here it is again enough to have the interchain hopping to generate the coupling. 

\section{Conclusions and Acknowledgements}

 In this paper I have shown that the models which combine conduction and localized electrons   provide a platform for very intricate types of order where the conduction electrons bind to slow collective modes of the spin subsystem. As a result the localized spins and the conduction electrons together create spin liquids with gapped fractionalized excitations. The local order parameters (COPs) include bound states of more than two electrons and are not amenable to analysis based on perturbative methods.  The discussion has rotated around quasi-one-dimensional models (the ones I dubbed Kondo-Heisenberg Cable arrays) where these fractionalized excitations remain one-dimensional even when different KH cables are coupled in $D>1$ array. 
 
  Composite orders naturally give rise to rich order parameter manifolds which include various types of density waves, including those of pairs and quartets of electrons. The formation of the spin liquid is accompanied  by a simultaneous formation of the order parameter  amplitudes, but the phase coherence is established only by three-dimensional interactions in the cable array. As a consequence the   magnitudes of the transition temperatures are not related to the spin gaps. 
  
   If the electron hopping matrix elements between different cables exceed the spin gap, pockets of quasiparticle Fermi surface appear. As it has been pointed out in \cite{FL}, KHC model reproduces many features found in the pseudogap phase of the cuprates. 
 
  It remains to be seen whether  the present ideas can be generalized for isotropic models in $D>1$. As we know from the literature on spin liquids, to propagate in $D>1$ dimensions fractional particles need to have companions in the form of visons. For instance, in the exactly solvable  Kitaev model \cite{kitaev} of spin liquid the role of visons for  propagating Majorana fermions is played by  static Z$_2$ gauge field fluxes. They facilitate a propagation of  the Majorana fermions in all lattice directions. As far as I can see there are no visons in the present construction and the fractional particles remain one dimensional. 


 I am grateful to E. Fradkin for asking the right question, for C. Chamon, L. Fidkowski, G. Kotliar, P. Lecheminant, G. Mussardo and  T. M. Rice  for interesting discussions. The work was supported by Center for Computational Design of Functional Strongly Correlated Materials and Theoretical Spectroscopy under DOE grant DE-FOA-0001276.

\newpage

\widetext

\appendix

\section{Exact solution of the WZNW model perturbed by the current-current interaction}

 The exact solution of the WZNW model perturbed by the current-current interaction can be derived from the relativistic limit of the fermion model 
\bea
H = \int \rd x \Big(-\frac{1}{2m}\psi^+_{j\alpha}\p_x^2\psi_{j\alpha}  - \mu\psi^+_{j\alpha}\psi_{j\alpha} - U\psi^+_{j\alpha}\psi^+_{j\beta}\psi_{i\beta}\psi_{i\alpha}\Big),
\eea
where $j,i = 1,...n; \alpha,\beta = 1, 2$ and $U>0$. 
This model is exactly solvable by the Bethe ansatz \cite{me}. The relativistic limit is obtained by the spectrum linearization (\ref{ep},\ref{psi}). The interaction then becomes 
\bea
- U\psi^+_{j\alpha}\psi^+_{j\beta}\psi_{i\beta}\psi_{i\alpha} \rightarrow U{\bf F}_R{\bf F}_L - UJ^a_RJ^a_L,
\eea
where $F^a_R,F^a_L$ are SU$_n$(2) and $J^a_R,J^a_L$ ($a=1,... n^2-1$) are SU$_2$(n) currents. The latter interaction is marginally irrelevant and can be discarded. As a result the only gapped sector is the one described by the SU$_n$(2) WZNW perturbed by the current-current interaction. 

 The solution can be also extracted from the following  Bethe ansatz equations: 
 \bea
 && [e_n(u_a -v_F/J_K)]^L[e_n(u_a +v_H/J_K)]^L = \prod_{b=1}^Me_2(u_a-u_b), ~~ S^z=nL/2 -M,\nonumber\\
 && E = \frac{1}{2\ri}\sum_a\Big[v_F\ln e_{n}(u_a -v_F/J_K) - v_H\ln e_{n}(u_a + v_H/J_K)\Big]
\eea
where 
\[
e_j(x) = \frac{x-\ri j/2}{x+\ri j/2}
\]
The Thermodynamic Bethe ansatz equations in the continuum limit are \cite{me}: 
\bea
&& \epsilon_j(\theta) = Ts*\ln[1 + \re^{\epsilon_{j-1}(\theta)/T}][1 + \re^{\epsilon_{j+1}(\theta)/T}] - \delta_{j,n} E(\theta), ~~ j=1,2,...\\
&& F/L = E_0 -T\int \frac{\rd P(\theta)}{2\pi} \ln[1+\re^{\epsilon_n(\theta)/T}], \\
&& s*f(x) = \int \frac{\rd y}{\pi \cosh(x-y)}f(y)
\eea
where $E(\theta), P(\theta)$ are given by Eq.(\ref{param}).  Expanding at $T << \Delta$ one obtains 
\bea
F/L = - T{\cal Q}\int \frac{\rd P}{2\pi}\re^{- E(\theta)/T}, ~~{\cal Q} = 2\cos\Big(\frac{\pi}{n+2}\Big), 
\eea
with ${\cal Q}$ being the so-called quantum dimension. This number indicates that the state of $N >> 1$ particles with energy $E$ is degenerate and the degeneracy is approximately ${\cal Q}^N$. The fact that ${\cal Q}$ is not an integer is an indication of the non-Abelian nature of the excitations.

\section{Useful facts about $n=2$ KHC model.}

 In classification of fermionic fields of the $n=2$ 1DEG I follow the scheme described in \cite{maldacena}. Namely, one introduce four bosonic holomorphic fields 
 \be
 \varphi_c, ~~\varphi_f, ~~ \varphi_s, ~~ \varphi_{sf},
 \ee
 with their antiholomorphic counterparts  $\bar\varphi_a$ ($a=c,f,s,sf$) to bosonize the fermions:
 \bea
 && R_{p,\s} = \frac{\lambda_{p\s}}{\sqrt{2\pi a_0}}\exp[\ri\sqrt{\pi}(\varphi_c +p\varphi_f + \s\varphi_s +p\s\varphi_{sf})], \nonumber\\
 && L_{p,\s} = \frac{\lambda_{p\s}}{\sqrt{2\pi a_0}}\exp[-\ri\sqrt{\pi}(\bar\varphi_c +p\bar\varphi_f + \s\bar\varphi_s +p\s\bar\varphi_{sf})],
 \eea
 where $\lambda_{p\s}$ are anticommuting Klein factors. Then right-moving the Majorana fermions are 
 \bea
 && \eta_1 = \frac{\xi_c}{\sqrt{2\pi a_0}}\cos(\sqrt{4\pi}\varphi_c), ~~ \eta_2 = \frac{\xi_c}{\sqrt{2\pi a_0}}\sin(\sqrt{4\pi}\varphi_c), \nonumber\\
 && \eta_3 = \frac{\xi_f}{\sqrt{2\pi a_0}}\cos(\sqrt{4\pi}\varphi_f), ~~ \eta_4 = \frac{\xi_f}{\sqrt{2\pi a_0}}\sin(\sqrt{4\pi}\varphi_f), \nonumber\\
 && \eta_5 = \frac{\xi_{sf}}{\sqrt{2\pi a_0}}\cos(\sqrt{4\pi}\varphi_{sf}),
 \eea
 where $\xi_a$ are anticommuting Klein factors and 
 \bea
 \chi_1 = \frac{\xi_s}{\sqrt{2\pi a_0}}\cos(\sqrt{4\pi}\varphi_s), ~~ \chi_2 = \frac{\xi_s}{\sqrt{2\pi a_0}}\sin(\sqrt{4\pi}\varphi_s), ~~ \chi_3 = \frac{\xi_{sf}}{\sqrt{2\pi a_0}}\sin(\sqrt{4\pi}\varphi_{sf})
 \eea
 It is assumed that the bosonic fields are governed by the Gaussian action.
 
 The Ising order and disorder parameters are related to $\Phi = \varphi +\bar\varphi, ~~ \Theta = \varphi -\bar\varphi$ fields. If one takes two copies of the critical Ising model we have \cite{zuber}
 \bea
 && \s_1\s_2 = \frac{1}{(\pi a_0)^{1/4}}\sin(\sqrt{\pi}\Phi), ~~\mu_1\mu_2 = \frac{1}{(\pi a_0)^{1/4}}\cos(\sqrt{\pi}\Phi), \nonumber\\
 && \s_1\mu_2 = \frac{1}{(\pi a_0)^{1/4}}\sin(\sqrt{\pi}\Theta), ~~\mu_1\s_2 = \frac{1}{(\pi a_0)^{1/4}}\cos(\sqrt{\pi}\Theta). \label{zuber}
 \eea
 
   The most convenient and economic way to establish a correspondence between different representations of the 2-leg problem is to use the SU$^s_2$(2)$\times$SU$^f_2$(2) basis and employ the non-Abelian bosonization. One SU(2) group represents rotations generated by currents of total spin and the other by chain currents. Transition from chain to band representation can be viewed as a rotation basis in  SU$^f(2)$ space. To make sure this approach is sound I will make a cross-check with the Abelian bosonization. 
  
   There is one subtlety discussed in \cite{QFT}. Namely, the group we are dealing with is not really SU(2), but its complexification SU(2,C).

 Below there are examples of OPs which are spin singlets. Being fused with the spin matrix $h$ they will leave 
 \be
 \re^{\pm i\sqrt{\pi}\Phi}G, ~~ \re^{\pm i\sqrt{\pi}\Theta}G
 \ee
 as the fluctuating COPs. All this can be expressed as products of 5 Ising fields which constitutes the 16-dimensional spinor representation of the SO(5) group. An important qualitative difference with the single chain case is that the COPs include pairs with momentum $\neq \pi$. 
    
  Using the standard bosonization rules  I derived the following formulae.
  
  The $s$-wave CDW order parameter. In the chain representation we have 
  \bea
  && R^+_{1\s}L_{1\s} + R^+_{2\s}L_{2\s} = 2\ri \re^{\ri\sqrt\pi \Phi_c}\Big[\re^{\ri\sqrt\pi\Phi_f}\cos\sqrt\pi(\Phi_s + \Phi_{sf}) + \re^{-\ri\sqrt\pi\Phi_f}\cos\sqrt\pi(\Phi_s - \Phi_{sf})\Big]\nonumber\\
  && 4\ri \re^{\ri\sqrt\pi\Phi_c}\Big(\cos\sqrt\pi\Phi_f \cos\sqrt\pi\Phi_s\cos\sqrt\pi\Phi_{sf} + \ri \sin\sqrt\pi\Phi_f \sin\sqrt\pi\Phi_s\sin\sqrt\pi\Phi_{sf}\Big) = \nonumber\\
  && -4 \re^{\ri\sqrt\pi\Phi_c}\Big(M_{1}M_2 M_3 \mu_1\mu_2\mu_3 + \ri \Sigma_1\Sigma_2\Sigma_3\s_1\s_2\s_3\Big) = \nonumber\\
  &&-\frac{1}{4}\re^{\ri\sqrt\pi\Phi_c}\Big[\mbox{Tr}(G + G^+)\mbox{Tr}(g + g^+) - \ri \mbox{Tr}(G - G^+)\mbox{Tr}(g - g^+)\Big].\label{sCDW}
  \eea
  Here $M,\mu$ are disorder and $\Sigma,\s$ order parameter fields of the Ising models describing the flavor and spin sectors, $G$ and $g$ are matrices from the flavor and spin sector respectively.  In the band representation the expression in terms of fermions looks the same with chain indices $1,2$ being replaced by band indices $a,b$. Naturally the expression in terms of matrices looks the same, as it should be. 
  
   Now let us consider a more general CDW OP:
   \bea
   {\cal O}^a_{CDW} = R^+_{j\s}\tau^a_{jk}L_{k\s}.
   \eea
   The simplest member of this family is 
   \bea
   && {\cal O}^3 =  2\ri \re^{\ri\sqrt\pi \Phi_c}\Big[\re^{\ri\sqrt\pi\Phi_f}\cos\sqrt\pi(\Phi_s + \Phi_{sf}) - \re^{-\ri\sqrt\pi\Phi_f}\cos\sqrt\pi(\Phi_s - \Phi_{sf})\Big] = \nonumber\\
   && 4\re^{\ri\sqrt\pi\Phi_c}\Big(-\sin\sqrt\pi\Phi_f\cos\sqrt\pi\Phi_s\cos\sqrt\pi\Phi_{sf} - \ri \cos\sqrt\pi\Phi_f\sin\sqrt\pi\Phi_s\sin\sqrt\pi\Phi_{sf} \Big) = \nonumber\\
   && -4\re^{\ri\sqrt\pi\Phi_c}\Big(\Sigma_1\Sigma_2 M_3\mu_1\mu_2\mu_3 + \ri M_1M_2\Sigma_3 \s_1\s_2\s_3\Big)= \nonumber\\
   && \frac{1}{4}\re^{\ri\sqrt\pi\Phi_c}\Big\{\mbox{Tr}(g-g^+)\mbox{Tr}[\tau^3(G-G^+)] +\ri \mbox{Tr}(g + g^+)\mbox{Tr}[\tau^3(G+G^+)]\Big\}.
   \eea
   
   The superconducting SCd ($\lambda_{2\uparrow}\lambda_{1\downarrow} = \lambda_{1\uparrow}\lambda_{2\downarrow} = \ri)$:
   \bea
   && \Delta_d = R_{1\uparrow}L_{2\downarrow} + R_{2\uparrow}L_{1\downarrow} - (\uparrow \rightarrow \downarrow) = \nonumber\\
   && \lambda_{1\uparrow}\lambda_{2\downarrow}\Big[\re^{\ri\sqrt{4\pi}(\varphi_{1\uparrow} - \bar\varphi_{2\downarrow})} + \re^{\ri\sqrt{4\pi}(\varphi_{2\uparrow} 
   - \bar\varphi_{1\downarrow})} \Big] + \lambda_{2\uparrow}\lambda_{1\downarrow}\Big[\re^{\ri\sqrt{4\pi}(\varphi_{1\downarrow} - \bar\varphi_{2\uparrow})} + \re^{\ri\sqrt{4\pi}(\varphi_{2\downarrow} - \bar\varphi_{1\uparrow})} \Big] =\nonumber\\
   && 2\re^{\ri\sqrt\pi\Theta_c}\Big[\re^{\ri\sqrt\pi\Phi_f}\cos\sqrt\pi(\Phi_s + \Theta_{sf}) + \re^{-\ri\sqrt\pi\Phi_f}\cos\sqrt\pi(\Phi_s - \Theta_{sf})\Big] = \nonumber\\
   && 4\re^{\ri\sqrt\pi\Theta_c}\Big(M_1M_2\Sigma_3\mu_1\mu_2\mu_3 + \ri \Sigma_1\Sigma_2M_3\s_1\s_2\s_3\Big)= \nonumber\\
   && \frac{1}{4}\re^{\ri\sqrt\pi\Theta_c}\Big\{\mbox{Tr}(g + g^+)\mbox{Tr}[\tau^3(G - G^+)] +\ri \mbox{Tr}(g - g^+)\mbox{Tr}[\tau^3(G + G^+)] \Big\}.
   \eea

\section{ The detailed description of the composite OPs}

 In this Appendix I discuss the formation of the simplest COP (\ref{cop}) for the case $n=2$.  As the first step of the proof I recast the products of the OPs of the 1DEGs and the TBC in terms of the operators of the GN models (\ref{Rl},\ref{Lr}). More precisely, we have to express the order and disorder parameters of the band fermions and the TBC antiferromagnet (they carry labels $F$ and $H$ respectively) in terms of  the corresponding operators of models (\ref{Rl},\ref{Lr}) labeled $R$ and $L$. I will use the Abelian bosonization formulae (\ref{zuber}).  Consider, for instance the product 
 \bea
 && (\s_1\s_2)_F(\s_1\s_2)_H \sim 2\sin(\sqrt\pi\Phi_F)\sin(\sqrt\pi\Phi_H) = \cos[\sqrt{\pi}(\varphi_F +\bar\varphi_F - \varphi_H -\bar\varphi_H)] - \cos[\sqrt{\pi}(\varphi_F +\bar\varphi_F + \varphi_H +\bar\varphi_H)] = \nonumber\\
 && \cos[\sqrt\pi(\Theta_L - \Theta_R)] - \cos[\sqrt\pi(\Phi_L +\Phi_R)] = (\mu_1\s_2)_L(\mu_1\s_2)_R - (\s_1\mu_2)_L(\s_1\mu_2)_R - (\mu_1\mu_2)_L(\mu_1\mu_2)_R - \nonumber\\
 && (\s_1\s_2)_L(\s_1\s_2)_R.
\eea
Hence it is plausible that the product of $F$ and $H$ OPs contains products 
\be
(\mu_1\mu_2\mu_3)_L(\mu_1\mu_2\mu_3)_R, ~~ (\s_1\s_2\s_3)_L(\s_1\s_2\s_3)_R
\ee
Such products have nonzero expectation values at least in some of the degenerate vacua of (\ref{GN1},\ref{GN2}). These expectation values may have a different sign in different vacua, but this does not affect the correlation functions of the COPs, since the two-point functions contains only squares of the amplitudes:
\bea
\la 0_j| (\s_1\s_2\s_3)_L(1)(\s_1\s_2\s_3)_L(2)|0_j\ra = [\la 0_j|(\s_1\s_2\s_3)_L(0)|0_j\ra]^2.
\eea


\end{document}